\documentclass[noacm,sigconf]{acmart}
\renewcommand\footnotetextcopyrightpermission[1]{} %

\AtBeginDocument{%
  }

\usepackage{multirow}
\usepackage{graphicx}
\usepackage{caption}
\usepackage{url}
\usepackage{makecell}

\usepackage{subcaption}
\usepackage{booktabs}

\usepackage{listings}

\lstdefinelanguage{Julia}{
  morekeywords={using,function, end, for, if, else, while, return},
  sensitive=true,
  morecomment=[l]\#,
  morestring=[b]"
}

\usepackage{etoolbox}
\usepackage{float}
\floatstyle{ruled}
\newfloat{code}{thp}{lop}
\floatname{code}{Code Block}

\usepackage{caption}
\begin{document}

\title[]{ElectricityEmissions.jl: A Framework for the Comparison of Carbon Intensity Signals}
\pagestyle{plain}

\author{Joseph Gorka, Noah Rhodes, and Line Roald}
\email{{ gorka, nrhodes, roald }@wisc.edu}
\affiliation{%
  \institution{University of Wisconsin -- Madison}
  \city{Madison}
  \state{Wisconsin}
  \country{USA}
}

\renewcommand{\shortauthors}{Gorka, Rhodes, and Roald}

\begin{abstract}
An increasing number of individuals, companies and organizations are interested in computing and minimizing the carbon emissions associated with their real-time electricity consumption. To achieve this, they require a carbon signal, i.e. a metric that defines the real-time carbon intensity of their electricity supply. Unfortunately, in a grid with multiple generation sources and multiple consumers, the physics of the system do not provide an unambiguous way to trace electricity from source to sink. As a result, there are a multitude of proposed carbon signals, each of which has a distinct set of properties and method of calculation.  It remains unclear which signal best quantifies the carbon footprint of electricity. This paper seeks to inform the discussion about which carbon signal is better or more suitable for two important use cases, namely carbon-informed load shifting and carbon accounting. We do this by developing a new software package ElectricityEmissions.jl, that computes several established and newly proposed carbon emission metrics for standard electric grid test cases. We also demonstrate how the package can be used to investigate the effects of using these metrics to guide load shifting.
Our results affirm previous research, which showed that the choice of carbon emission metric has significant impact on shifting results and associated carbon emission reductions. 
In addition, we demonstrate the impact of load shifting on both the consumers that perform the shifting and consumers that do not. Disconcertingly, we observe that shifting according to common metrics such as average carbon emissions can \emph{reduce} the amount of emissions allocated to the consumer doing the shifting, while \emph{increasing} the total emissions of the power system.

\end{abstract}

\maketitle

\section{Introduction}
As the impacts of climate change become increasingly visible, many electricity consumers, from private persons to large corporations, have shown increasing interest in assessing and reducing their carbon footprint. For many consumers, a significant portion of their carbon footprint is tied to their electricity consumption and associated emissions from electricity generation. The source of electricity generation (e.g. solar, wind, hydro, nuclear, natural gas or coal) as well as the temporal availability of renewable power (i.e. daily and seasonal cycles of solar, wind and hydropower) vary widely between locations, leading to differences in carbon intensity of electric power across time and space. This provides an opportunity for consumers to reduce carbon emissions by using power where and when low-carbon electricity is available. However, to actively reduce the carbon emissions of their electricity supply, consumers need %
real-time information about how carbon emissions of electricity vary. %
Furthermore, they need a framework for leveraging this information not only for real-time load shifting but also for long-term carbon accounting, such that they can get credit for their efforts to reduce emissions. This is particularly important for companies and industries that are subject to emerging carbon emission regulations and reporting requirements, including e.g. producers of clean hydrogen \cite{proposed-rule}.

Quantifying the carbon intensity of electricity consumption in a grid with multiple consumers and multiple sources of generation, is however not straightforward. While it is possible to obtain information on emissions from the generators and assess how the total carbon footprint varies over time, the physics of the grid do not lend themselves to a clear definition of the generation mix (and hence emissions intensity) that serves individual locations. This leaves significant open questions regarding which carbon emissions is the best, see e.g. \cite{sukprasert2024implications, RICKS2024114119}. This paper develops a software package and an analysis framework to help evaluate the effectiveness of different carbon emission metrics.

\subsection{Related work}
Existing research has focused on computing and analyzing different carbon emission metrics for real-time electricity consumption, e.g. in \cite{electricitymaps, callaway2018location, wattime, lindberg2020environmental, deka2023contributions, chen2023carbon}, spurring an interest in how consumers might use such emissions metrics to assess and minimize the carbon impact of their electricity consumption. Previous studies have examined the potential to reduce electricity emissions in the context of data-centers \cite{Lin2012,lindberg2020environmental,chien2020,Radovanovic2021}, production of hydrogen fuel \cite{Ricks2022Minimizing,Zhang2020Flexible_hydro}, and residential consumption \cite{Mata2020A,Harris2015Residential}. The increasing availability of real-world carbon emissions data, provided both directly by grid operators \cite{pjm_lmce, carbonintensity} and by third party organizations \cite{wattime,electricitymaps}, has enabled real-world implementations of load shifting strategies guided by emission metrics\cite{Radovanovic2021,Maji2023Bringing,apple_clean_energy_charging, greenlight}. For example, battery-powered devices such as phones have settings that enable ``clean energy charging'' \cite{apple_clean_energy_charging}, while customers in UK can install light bulbs that turn green when the electricity is green \cite{greenlight}.

Generally, existing works have tended to focus on a single carbon metric--e.g. to analyze its properties or use it as input to a theoretical or realized emissions-reduction scheme. Recently, however, there has also been increased interest in \emph{comparing} different metrics for the purposes of carbon accounting and informing load-shifting applications \cite{lindberg2021guide,lindberg2022using,wattimeblog2023marginal,electricitymapsblog2023marginal}. 
Two main shifting metrics that are currently in use. `Average Carbon Emissions' (ACE), defined as total system emissions divided by total system load, is the carbon metric used by the Greenhouse Gas Protocol for so-called location-based carbon accounting \cite{ghg2015}. `Locational Marginal Carbon Emissions' (LMCE), defined as the change in emissions due to a small change in load, %
has been investigated in a number of research studies and is available from at least one grid operator. %
A variety of works have examined the effectiveness of ACE and LMCE for informing load-shifting applications \cite{lindberg2021guide,lindberg2022using,sukprasert2024implications}. In \cite{lindberg2021guide}, a synthetic grid case study is used to demonstrate that the emissions impact of data-center load shifting based on either ACE or LMCE (as well as alternate data such as price and renewable curtailment) has a significant impact on shifting outcomes.
In a recent paper, the authors of \cite{sukprasert2024implications} compare real-world marginal and average carbon emissions signals from 65 regions throughout the world, and found the two signals to be largely negatively-correlated, leading to significant differences in load shifting outcomes and accounted emissions.

The debate between ACE and LMCE, and resulting illumination of their respective shortcomings, has also led to interest in defining new classes of carbon metrics. One notable example of this are so-called `carbon-flow'-based metrics \cite{kang2015carbon,chen2023towards,deka2023contributions}, which employ various non-physical assumptions to trace power flow (and associated carbon emissions) from generators to loads. Though not yet extensively analyzed, the method by which these carbon flow metrics are calculated is distinct from both ACE and LMCE. As a result, such metrics likely represent yet more competing definitions of carbon intensity, highlighting the increasing need to study the pros and cons of each type of metric.

Inter-metric variability in shifting-incentive and emissions accounting properties, such as that demonstrated in \cite{lindberg2021guide} and \cite{sukprasert2024implications}, highlights the need for new tools and analysis frameworks to compare different emissions metrics. This is particularly true given the high likelihood that new metrics will continue to be proposed and employed to inform consumers' beliefs regarding their carbon footprint, as well as their load-shifting behavior.

To inform the debate about the effectiveness and appropriateness of different carbon emission metrics, we develop a software package for comparing the different metrics from a power system perspective. 
Our comparison framework is based on modeling the US market clearing mechanism of an electric grid (represented by solving a DC optimal power flow (OPF) to determine the optimal cost-minimizing generator set-points) and computing total carbon emissions for the overall grid as well as various carbon emission metrics for different locations based on the result. This information can then be used for both carbon accounting (i.e. to allocate carbon emissions to different loads) and as an input to simulate load-shifting, where a carbon-sensitive load, such as a data center, changes their consumption in response to the carbon metrics. Finally, it is possible to assess the effectiveness of load shifting by re-solving the electricity market clearing with the updated electricity consumption pattern.

One important benefit of the framework is that it allows us (and other researchers) to evaluate both whether a given metric is effective in guiding load shifting (i.e. if a consumer reduces their use of electricity in an hour with a high emission value and increases it in an hour with low emission value, does this lead to a reduction in overall system emissions?) and whether it possesses desirable properties for carbon accounting (e.g. does the metric guarantee that the total carbon emissions assigned to consumers are equal to the total carbon emissions from generation?). 
This can help inform the debate around current metrics. However, our second goal is to provide a platform for evaluation of newly proposed metrics. We demonstrate this by proposing a new carbon emission metric, namely the adjusted locational marginal emissions (ALMCE) metric, and assessing its merits relative to existing metrics using our framework.

\subsection{Contributions}
\begin{itemize}
    \item We provide a qualitative overview of existing carbon emission metrics and their pros and cons, and propose the new ALMCE metric to try to combine benefits of several metrics.
    \item We contribute a new open source Julia package, "ElectricityEmissions.jl" that enables easy calculation of several existing and emerging carbon emissions metrics, including the newly proposed ALMCE metric.
    \item We demonstrate how the package can support quantitative comparison of different carbon emission metrics, both with and without load shifting, in a case study on the standard RTS-GMLC test system. 
    \item This case study is, to the best of our knowledge, the first to assess the impact of load shifting not only on total carbon emissions, but also on the accounted carbon emissions that are allocated to different types of loads.
\end{itemize}

\section{Background}
To explain the foundations of the real-time carbon accounting for electricity consumption, we start with a brief overview of the prevalent electricity market design in the United States, the emissions from generators and how changes in electricity consumption can impact these emissions.

\paragraph{Notation} In the following discussions, we will denote quantities related to electricity consumption by subscript $D$ and the set of loads by $\mathcal{D}$, while quantities associated with electricity generation are denoted by subscript $G$ and the set of generators is denoted by $\mathcal{G}$. We will generally use lower case $e$ to refer to carbon emissions intensities (measured in [TonsCO2/MW]) and capital $E$ to refer to carbon emissions (measured in [TonsCO2]). Lower case $p$ will be used to denote
quantities of power generation or consumption.

\paragraph{\textbf{Electricity Market Clearing}} We consider a region where the electricity market clearing can be modeled mathematically as an optimal power flow (OPF) problem. This is representative of an electricity market based on nodal pricing, which is the prevalent market design in the United States. %
In this market design, generators provide offers to generate electricity, which are associated with a given cost [\$/MWh] and capacity [MW]. The electricity consumption of electric loads is typically forecasted by the utilities that serve those loads, and the consumption is treated as a fixed value. This modeling is appropriate for most electric loads, which buy electricity on fixed rates through their local utility. However, the OPF model could easily be extended to also consider flexible, price-sensitive loads that source electricity directly from the whole sale energy market by modeling them as negative generation.

Given that the load is assumed to be fixed, the goal of the market clearing, and the objective of the corresponding OPF problem, is to minimize the total cost of generating electricity. This cost minimization is subject to constraints that model the physical flows of electric power and limits on generation and transmission capacity. It can be expressed as a linear optimization problem. We provide the mathematical formulation and additional explanations for this optimization problem in Appendix \ref{app:dcopf_formulation}.  %

Electricity markets clear in two periods, day-ahead (DA) and real-time (RT).  The DA market, solves a multi-time period optimization problem for the generation at each hour of the following day.  The RT market is solved every 5 minutes during the day, where discrepancies from the load forecast or renewable energy forecast used in the DA market are resolved by solving an OPF and adjusting the generator output.  
For either market, the primary outcome of the OPF is the optimal generation dispatch, given by a set of generation set-points $p_{G,i,t}^*$ for all generators $i\in \mathcal{G}$ at time step $t$, that is needed to satisfy system load $p_{D,j,t}$, representing the demand by loads $j\in \mathcal{D}$ at time $t$.

\paragraph{\textbf{Emissions from generators} } Based on the optimal generation dispatch $p_{G,t}^*$, we compute the carbon emissions as 
\begin{align}
    E^{tot}_{G,t} = \sum_{i\in\mathcal{G}} E_{G,i,t}(p_{G,i,t}^*)=e_{G,i,t} p_{G,i,t}^*
\end{align}
where $E_{G,i}(p_{G,i,t}^*)$ is a function describing the emissions associated with generating the power $p_{G,i,t}^*$ from generator $i$. This function could take many forms (for example, generators may have different efficiency levels depending on how much they produce or the ambient temperature), but for practical reasons and lack of more detailed data, we assume that the carbon emissions per MW of generated power is constant, such that $E_i(p_{G,i}^*)=e_{G,i} p_{G,i,t}^*$ where $e_{G,i}$ is the (constant) emissions factor of generator $i$. 

\paragraph{\textbf{Impact of changes in electricity consumption}} In the OPF problem, increasing or decreasing consumption $p_{D,t+1}$ will change the optimal generation set-points $p_{G,t+1}^*$ and the resulting carbon footprint of the grid. This allows consumers to impact carbon emissions by changing their consumption, for example by shifting their consumption to a different location (e.g. by shifting computing loads within networks of data centers) or to a different time step (e.g. by deferring consumption until later in the day). We note that while it is possible that a consumer might shift their load to better align with behind-the-meter renewable assets (hence lowering their emissions), the OPF-based framework presented in this work considers only consumer impact on utility-scale generation.

\section{Carbon Emission Metrics}
\label{sec:metrics}
To define the carbon emissions of electricity consumption, we need to define a time- and location-specific \emph{carbon emission metric} $e_{D,j,t}$ [TonsCO2/MW] that defines the carbon intensity per unit of electricity consumed at each timestep and location in the grid. Given the carbon emission metric $e_{D,j,t}$, the total carbon emissions of demand $j$ at time $t$ are given by
\begin{equation}
    E_{D,j,t} = e_{D,j,t} p_{D,j,t}
\end{equation}
where $p_{D,j,t}$ [MW] denotes the total electricity consumption and $E_{D,j,t}$ [TonsCO2] denotes the total emissions. 
The challenge lies in choosing how to define $e_{D,j,t}$ (often also referred to as carbon intensity or emissions factor). Even though the carbon emission factor from each electric generator $i$ and time step $t$, denoted by $e_{G,i,t}$, are well defined, there is no clear consensus about how to allocate these carbon emissions to different consumers. This is in part because there is no unambiguous method to trace the actual flow of electricity from generator to load. %
Thus, multiple definitions of carbon emission metrics have been proposed. We summarize the definitions of some widely used metrics below, along with a more recently proposed metric and a brand new metric that we consider here for the first time. 
Note that this is not intended to be an exhaustive list of carbon metrics. Some other metrics, such as long-run marginal emissions \cite{hawkes2014long}, are not included below as they are less relevant to real-time carbon accounting and load shifting.

\paragraph{\textbf{Average carbon emissions (ACE)}} The most intuitive and commonly used metric is \emph{average carbon emissions} $e_{D,t}^{avg}$, which is defined as the total carbon emissions incurred by the generators $E_{G,t}^{tot}$ and divided by the total power consumption\footnote{Note that care must be taken to appropropriately count the total carbon emissions from generation and the total load in cases where a region is importing or exporting electricity.}, 
\begin{equation}
    e_{D,t}^{avg} = \textstyle{\frac{E^{tot}_{G,t}}{\sum_{j\in\mathcal{D}} p_{D,j,t}}}.
\end{equation}
With this definition, the emissions factor $e_{D,j,t}=e_{D,t}^{avg}$ is the same for all loads $j$ in the grid at time $t$. 
As a metric, average carbon emissions has several benefits.
It is easily accessible (i.e. published by system operators or can be estimated from publicly available information without requiring a model of the network), is available for large geographical regions from data providers such as ElectricityMaps \cite{electricitymaps} and WattTime \cite{wattime}, and can be used for scope 2 electricity accounting under the Greenhouse Gas Protocol \cite{ghg2015}.
Average carbon emissions capture the emissions from all generators, and ensures that the total emissions from generation equal the total emissions assigned to loads. %
However, the average carbon emissions metric also has several drawbacks. It does not account for grid constraints that limit the transfer of power within a given region. %
Further, average carbon emissions does not accurately capture how 
load shifting, i.e. changes in the timing or location of electricity consumption of a specific load $p_{D,j,t}$, impacts total grid emissions. Another emerging drawback is that average carbon emissions does not reflect the availability of low-carbon generation sources that are currently not being used, and, in particular, the availability
of curtailed renewable energy. 

\begin{figure}%
    \includegraphics[width=0.45\textwidth]{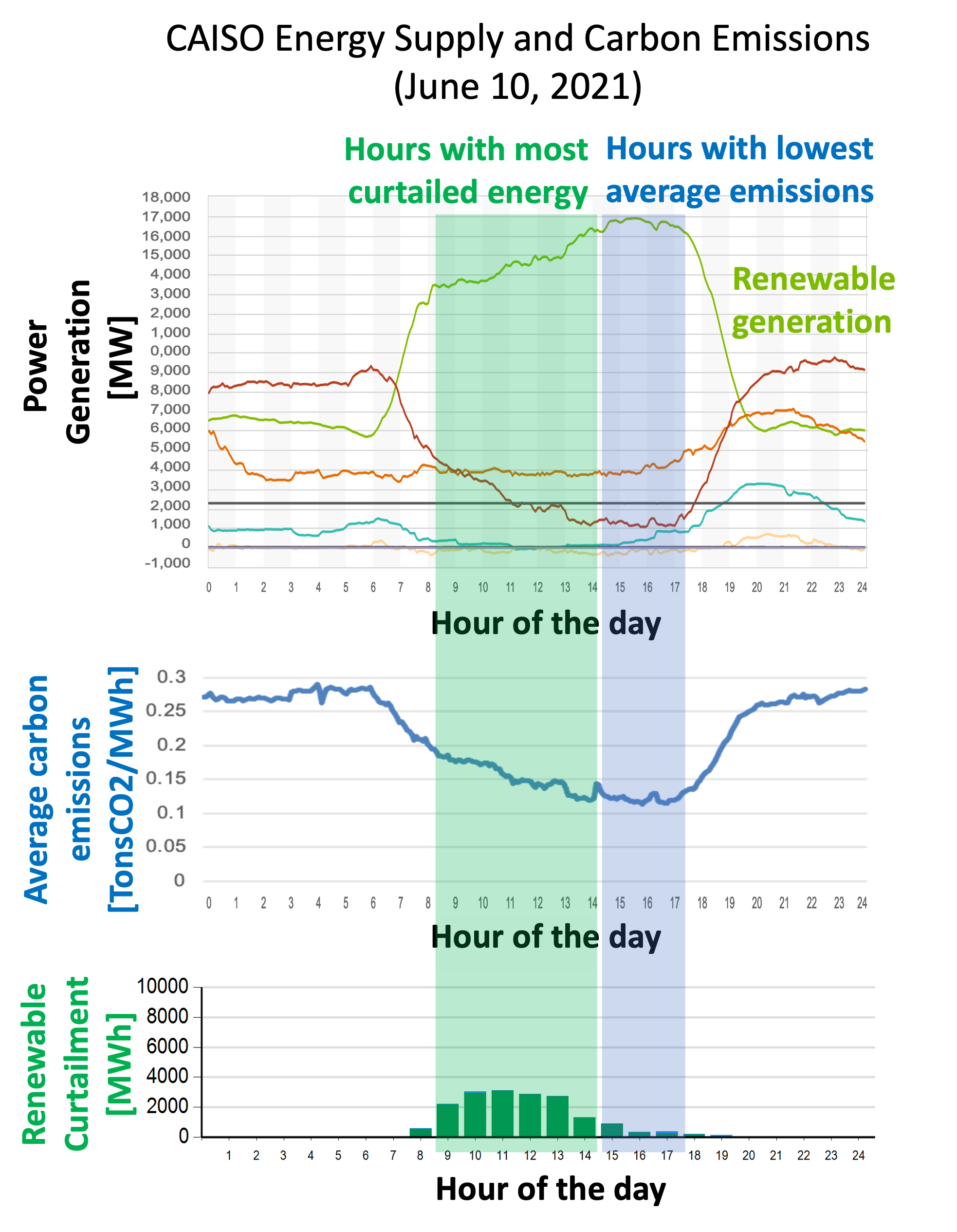}
 \caption{\emph{Power generation profiles (top), average carbon emission (middle) and renewable curtailment (bottom) for June 10, 2021 in CAISO. We highlight hours with high curtailment (green) and low average carbon emissions (blue). }}
 \label{fig:caiso}
\end{figure}

We illustrate these drawbacks based on an example from the CAISO grid (in California) on June 10, 2021. 
Fig. \ref{fig:caiso} shows the total generation for each fuel type (top), average carbon emissions measured in TonsCO2/MWh (middle) and amount of renewable curtailment in the grid (bottom). 
We observe that the average emissions are lowest in the late afternoon, when the renewable energy generation is highest. However, there is significant renewable energy curtailment earlier in the day, which is why the average emissions are higher in this time period. If the average emissions metric is used to shift load, consumers may choose to shift their power consumption until later in the day. However, from a grid perspective, this is opposite of what is needed, as such a shift would likely increase renewable energy curtailment and lead to overall higher emissions. %

\paragraph{\textbf{Locational marginal carbon emissions (LMCE)}} To answer questions such as “how do carbon emissions change if I choose to charge my electric vehicle now rather than later?”, a more accurate metric is the locational marginal carbon emissions.
By applying sensitivity analysis methods to the OPF problem, we can describe how a small change in a given load, denoted by $\delta p_{D,j,t}$, changes the optimal generation dispatch. We denote changes in generation by $\delta p_{G,j,t}^*$. Once we know $\delta p_{G,j,t}^*$, we can derive the associated change in grid emissions at time $t$ as $\delta E_{G,t}^{tot} = \sum_{i\in\mathcal{G}} e_{G,i,t} \delta p_{G,i,t}^*$. After normalizing for the change in load, we obtain the so called \emph{locational marginal carbon emissions} (LMCE)
\begin{equation}
    e_{D,j,t}^{lmce} = \textstyle{\frac{\delta E_{G,t}^{tot}}{\delta p_{D,j,t}}}.
    \label{eq:lmce_def}
\end{equation}
Our implementation of LMCE is based on the method described in \cite{lindberg2020environmental}, but extended to account for piece-wise linear cost functions. The full mathematical model is described in Appendix \ref{app:LMCE_calc}.  Note that LMCE inherently accounts for transmission constraints, and thus varies by location in the grid when transmission capacity is limited.

The locational marginal emissions $\lambda_{CO2}$ are arguably the metric that most accurately describes the impact of (small) load shifting actions (i.e. intentional changes in $p_{D,j,t}$) to reduce carbon emissions \cite{lindberg2021guide}, \cite{lindberg2022using}. 
However, because they represent sensitivities of the OPF solution, LMCE
can vary widely between time-steps and and may not be an appropriate metric for larger load shifts. 

Further, it is unclear how LMCE can be used for carbon accounting, as it
does not capture any information about emissions from non-marginal generators. 
Here, it is important to note the distinction between LMCE and locational marginal prices (LMPs), which are the result of clearing the electricity market. %
Since electricity markets seek to minimize total generation cost, the locational marginal price represents the marginal generation cost of the most expensive generator that is dispatched in the market clearing to serve electric load at a given location). The marginal generator can vary for different locations based on the network constraints, but all other dispatched generators are lower cost. For emissions, there is no such relationship for the marginal generator being the highest emitting dispatched generator. The marginal generator may have higher or lower emissions 
than the other units that are dispatched.
As a result, if we allocate carbon emissions to loads based on the locational marginal emissions as the emission intensity, the total allocated carbon emissions will generally be higher or lower than the total emissions from generation. In many cases we find that the difference between the total emissions from generation is significantly different from the total emissions allocated to loads.  %

\begin{figure}%
    \includegraphics[width=0.42\textwidth]{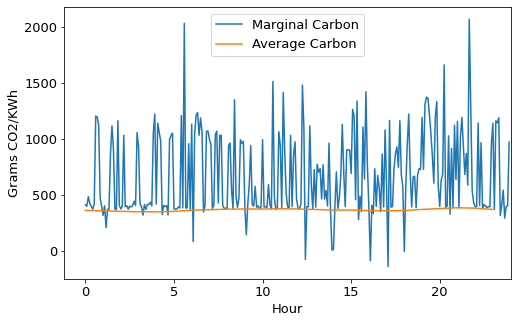}
 \caption{ \emph{Marginal (blue) and average (orange) carbon emissions for Chicago (data from \cite{pjm}).}}
 \label{fig:comed}
\end{figure}

To illustrate the volatility of LMCE, %
Fig. \ref{fig:comed} shows the average and locational marginal emissions for ComEd in Chicago on 8/13/23. This is real-world data reported by the utility. We observe that the 5 minute, real-time locational marginal emissions have much higher volatility than the hourly average emissions. We also see that the LMCE is often significantly higher than the average emissions, indicating that the marginal generator at this location is more polluting than the average generator.  

\paragraph{\textbf{Carbon flow-based metrics}} Another class of metrics that have been considered in the academic literature are metrics based on a ``carbon flow'' which ``traces'' carbon emissions from generators to load \cite{kang2015carbon, chen2023carbon, deka2023contributions}. The key idea behind these metrics is that generators inject both power and carbon emissions into the grid, and the carbon ``flows'' along with the electricity. The methods in \cite{kang2015carbon, chen2023carbon, deka2023contributions} are based on the so-called proportional power sharing rule \cite{bialek1996tracing, kirschen1999tracing}, an assumption which stipulates that the power flows into a bus (and the associated carbon emissions) are shared proportionally across the power flows out of it. It is important to note that this is an assumption, not a physical law. There exist a variety of other proposals for tracing power flow (see e.g. \cite{dhople2019tracing} and references therein), which would likely lead to different carbon allocations across loads.
To the knowledge of the authors, carbon flow metrics have not been adopted in practice, possibly because their definition is less intuitive and more mathematically involved. Nevertheless, these methods are interesting as they give rise to a range of possible carbon emission metrics, including locational average carbon emissions \cite{deka2023contributions}.

\paragraph{\textbf{Adjusted Locational Marginal Carbon Emissions (ALMCE)}}
Marginal metrics such as LMCE are appealing for use in informing load-shifting decisions, as they provide information on how grid emissions will change given a small change in load. However, performing carbon accounting using LMCE is problematic due to the lack of a guarantee that LMCE-assigned emissions will sum to actual total system emissions. To address this issue, 
we propose a new metric, the \emph{adjusted locational marginal carbon emissions} (ALMCE). The core idea of ALMCE is to augment LMCE with an adjustment factor, which serves to normalize the sum of assigned emissions to be equal to the actual system emissions total. %
Mathematically, ALMCE is defined for load $j$ at time $t$ as follows.
\begin{equation}
\label{eq:adjusted}
    e_{D,j,t}^{almce} = e_{D,j,t}^{lmce} + \frac{E_{G,t}^{tot}-E_{D,t}^{lmce}}{\sum_{j\in\mathcal{D}} p_{D,j,t}},
\end{equation}
where the first term is the locational marginal emissions $e_{D,j,t}^{lmce}$ and second term represents the adjustment. 
This adjustment is defined as the difference between the total emissions from generation $E_{G,t}^{tot}$ and the total emissions assigned to loads using the LMCE metric $E_{D,t}^{lmce}=\sum_{j\in\mathcal{D}} e_{D,j,t}^{lmce} p_{D,j,t}$, %
shared across all loads proportional to their consumption.

While the full properties of ALMCE for carbon accounting and load-shifting remain to be determined through further analysis empirical testing, there are some properties worth noting. First, ALMCE will fluctuate just as LMCE fluctuates, providing a rapid signal of the emissions of the marginal generators. Second,if there is a single marginal generator in the system, ALMCE is equivalent to ACE and in this case would lead to similar shifting results. Third, if we consider a single time step and sort nodes according to their nodal carbon emission values (e.g. from high to low), then ALMCE and LMCE would lead to the same ordering of nodes. Therefore, if we only consider spatial load shifting (where you can choose between consuming at different locations rather than at different times), ALMCE and LMCE might lead to similar shifting results. Finally, we note that the real-world implementation of ALMCE would not require significant effort in systems that already calculate LMCE, such as PJM \cite{pjm_lmce}, since the LMCE calculation is the most demanding part of computing ALMCE.

\section{ElectricityEmissions.jl}
\label{sec:EA.jl}
We have created a new Julia package, \emph{ElectricityEmissions.jl}
\footnote{https://github.com/WISPO-POP/ElectricityEmissions.jl}
, to serve as a central implementation of the carbon metrics mentioned in the previous section. The package allows the calculation of these metrics for power system test-cases in the popular MATPOWER format, opening the door for carbon-intensity analysis on thousands of simulated grids of varying sizes and properties.  \textit{ElectricityEmissions.jl} is open-source, and contributions of other carbon metrics and accounting methods are welcome. 
Below is an example of computing and plotting carbon metrics for a small five-bus network.

\begin{code}[h]
\caption{Calculating and plotting carbon intensity metrics}
\label{code:sim}
\begin{lstlisting}
using PowerModels, ElectricityEmissions
using HiGHS

# Load the network data with generator emissions intensity
network = PowerModels.parse_file("case5_gen_intensity.m")

# Calculate carbon intensities (solves a PWL DC OPF problem)
lmce  = calculate_LMCE(network, HiGHS.Optimizer)
almce = calculate_ALMCE(network, HiGHS.Optimizer)
ace   = calculate_ACE(network, HiGHS.Optimizer)
lace  = calculate_LACE(network, HiGHS.Optimizer)

# Calculate total system emissions
total_emissions = calculate_system_emissions(network, HiGHS.Optimizer)

# Plot network with carbon intensity data

# Update network data with chosen emissions intensity
update_emissions_intensity!(network,lmce)

# Plot the network
plot_emissions(network)
\end{lstlisting}
\end{code}

\subsection{Carbon Intensity Calculation}
The first part of the code example demonstrates how to compute the carbon intensity metrics. The MATPOWER-format test-case, containing network data and generator emissions information, is loaded using the \texttt{parse\_file} function from the \textit{PowerModels.jl} package \cite{powermodels_jl}. It is then passed to the functions \texttt{calculate\_ACE}, \texttt{calculate\_LMCE},  \texttt{calculate\_LACE} and
\texttt{calculate\_ALMCE} to calculate the respective carbon intensity metrics.  %
Each of these functions solve a DC-OPF problem to dispatch the system. Then load demand, optimal generator set-points, and generator carbon intensity (and for some metrics, the network structure) are used to calculate the nodal carbon intensity for each node in the network, as described in Section \ref{sec:metrics}. We calculate total system emissions with the function \texttt{calculate\_system\_emissions}, which only needs generator set-points as an input.

\subsection{Plotting}

Visualization of the carbon metrics is important to understand the geographic spread of emissions intensity.  %
To help users analyze their results, the functions 
\texttt{update\_emissions\_intensity!()} and
\texttt{plot\_emissions()} 
provide convenient data-processing and default settings for use with the power grid plotting package PowerPlots.jl \cite{powerplots}. 
An example of a visualization of the metrics is shown in the case study in Fig. \ref{fig:RTS_average}.

\subsection{Example Workflow}
\label{sec:workflow}
Here we step through an example workflow of using ElectricityEmissions.jl to estimate emissions and shift load, with counter-factual analysis. 

\textbf{\emph{0. Test-Case Setup}}
First, obtain a MATPOWER-format grid test-case. Test-cases with various properties (size, connectedness, distribution vs. transmission system, etc.), are included as part of the MATPOWER software package \cite{matpower} or are easily accessible via a library maintained by the IEEE Power and Energy Society \cite{pglib_opf}. Second, as shown in Code Block \ref{code:sim}, use the \textit{PowerModels.jl} \cite{powermodels_jl} package to load the test-case. Third, if not already included in the test-case, assign generator emissions intensities. \textit{ElectricityEmissions.jl} expects this information for each generator $i$ under the key "emissions", e.g. \texttt{network["gen"][i]["emissions"]}. Finally, it is best practice to ensure that no two generators at a given node have the same cost function, as such duplications can lead to problems calculating LMCE/ALMCE (see Appendix \ref{app:LMCE_calc} for details). This issue can be easily fixed by adding a small amount of noise to the offending generator cost functions.

 \textbf{\emph{1. Initial Emissions Calculations}} Prior to any load-shifting, use ElectricityEmissions.jl to calculate the initial carbon intensity at each bus in the system using the relevant \texttt{calculate\_...()} function for each of the carbon intensity metrics being compared. These intensity values will have units of \emph{Tons CO2/MWh}. To calculate assigned emissions (in Tons CO2) based on a particular metric, simply multiply the load at each bus by the corresponding carbon intensity value. Finally, store network intensities and assigned emissions according to each metric, as well as total system emissions (using \texttt{calculate\_system\_emissions}).

 \textbf{\emph{2. Solve a Load-Shifting Problem}} A load uses the carbon intensity values generated in the previous step (or outside carbon metric values/forecasts) as input to solve a load shifting optimization problem to minimize its emissions. Do this for each intensity metric and record the resulting post-shift loads. To \emph{estimate} the impact of this shifting on the accounted carbon emissions, multiplying the initial carbon intensities by the new load values. Note that these are just estimates, for two main reasons. First, the carbon intensities might change as a result of the load shifting, leading to differences between the estimated and realized accounted carbon emissions. Second, the carbon intensities are themselves estimates of the carbon footprint of electricity. Although a load shifting action leads to a significant reduction in the accounted carbon emissions, it is not guaranteed that the total system emissions would actually decrease by the same amount.  %

 \textbf{\emph{3. Post-Shift Emissions Calculation}}
Shifting load will result in new generator set-points and a new power flow which may change the carbon intensity values at each bus.  To determine the final post-shift emissions, we rerun the relevant \texttt{calculate\_...()} functions with the shifted load profile to recalculate the carbon intensity metrics. As explained above, the carbon intensities may change as a result of the load shifting.  Using the post-shift carbon intensities, it is possible to compute the accounted emissions that would be realized if the load-shifting was implemented. We refer to those as the \emph{realized} accounted emissions.

We can also calculate the total post-shift system emissions, and compare those with the initial system emissions to assess the effectiveness of the load shift in reducing total carbon emissions.

\section{Case Study: Accounting}
We first provide a case study on carbon accounting according to the different carbon emission metrics, leveraging the ElectricityEmissions.jl package for computations. %

\subsection{Dataset}
We make use of the RTS-GMLC \cite{RTSGMLC} test case, a popular synthetic dataset geographically ``located'' in the Southwestern United States. The system contains 73 buses, 120 lines, and 158 generators of varying types, including renewables (wind, solar, and hydro), natural gas, oil, and coal. 
One year's worth of hourly load and generation data is available for this system, totaling 8,784 cases. 
We use the day-ahead load and renewable energy data from the RTS-GMLC data to calculate these metrics on an hourly basis.
We make two modifications/additions, similar to the changes made to the same system in \cite{lindberg2021guide}. First, we add data-centers to the system by setting buses 103, 107, 204, and 322 to be data-centers, and refer to those data centers as DC 103, DC 107, DC 204 and DC 322, respectively. Each data center has a nominal load of 250MW, such that the total data-center load corresponds to slightly over 20\% of total energy consumption over the course of the year. Second, we use values from the U.S. Department of Energy\cite{environmentbaseline} to assign the following emissions intensities (in metric tons of CO2 per MWh) to each generator in the system:
\begin{itemize}
    \item Wind, Solar, Hydro: 0 
    \item Natural Gas: 0.6042
    \item Oil: 0.7434
    \item Coal: 0.9606
\end{itemize}
Note that the emission values in this table differ from the ones listed in \cite{lindberg2021guide} (where the emissions values for coal and natural gas appear to have been switched). The data and code are available as examples in the package documentation. 
\footnote{https://github.com/WISPO-POP/ElectricityEmissions.jl}

\subsection{Methodology}
We perform carbon accounting based on the LMCE, ALMCE, ACE, and LACE carbon intensity metrics across all time-steps in the dataset. ElectricityEmissions.jl is used to calculate nodal carbon intensities for all metrics, as well as total system emissions. 
Following this, accounted emissions are calculated for each metric through multiplication by nodal loads.

\subsection{Results}
\begin{figure*}%
    \centering
    \includegraphics[width=1\linewidth]{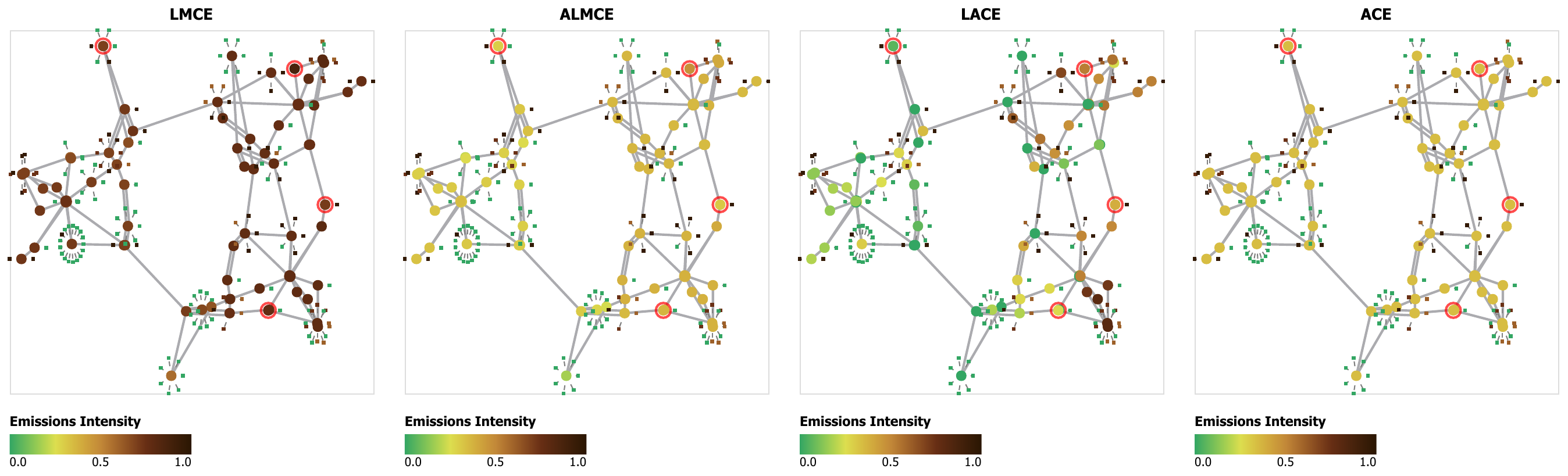}
    \caption{Annual average nodal carbon emissions intensity, by metric. The buses are shown as circles, and generators are small squares arrayed around the buses.  The color of the generators indicate the carbon emissions intensity of the source, while the color of the buses show the annual average carbon emissions. Locations of data centers are circled in red. Transmission lines connecting buses are shown in grey.} \label{fig:RTS_average}
\end{figure*}

\begin{table}[]
\resizebox{\columnwidth}{!}{%
\begin{tabular}{@{}rccccllclclcl@{}}
\multicolumn{1}{l}{}          & \textbf{Generated}           & \textbf{System} & \textbf{DC Total} & \multicolumn{3}{c}{\textbf{DC 103}} & \multicolumn{2}{c}{\textbf{DC 107}} & \multicolumn{2}{c}{\textbf{DC 204}} & \multicolumn{2}{c}{\textbf{DC 322}} \\ \cmidrule(l){2-13} 
\multicolumn{1}{l}{\textbf{}} & Sum                     & Sum             & Sum         & \multicolumn{3}{c}{Sum}             & \multicolumn{2}{c}{Sum}             & \multicolumn{2}{c}{Sum}             & \multicolumn{2}{c}{Sum}             \\ \midrule
\textbf{LMCE}                 & \multirow{4}{*}{15.828} & 33.012          & 6.692       & \multicolumn{3}{c}{1.686}           & \multicolumn{2}{c}{1.562}           & \multicolumn{2}{c}{1.917}           & \multicolumn{2}{c}{1.527}           \\
\textbf{ALMCE}                &                         & 15.828          & 3.162       & \multicolumn{3}{c}{0.803}           & \multicolumn{2}{c}{0.679}           & \multicolumn{2}{c}{1.035}           & \multicolumn{2}{c}{0.644}           \\
\textbf{ACE}                  &                         & 15.828          & 3.008       & \multicolumn{3}{c}{0.752}           & \multicolumn{2}{c}{0.752}           & \multicolumn{2}{c}{0.752}           & \multicolumn{2}{c}{0.752}           \\
\textbf{LACE}                 &                         & 15.828          & 2.707       & \multicolumn{3}{c}{0.577}           & \multicolumn{2}{c}{0.850}           & \multicolumn{2}{c}{1.153}           & \multicolumn{2}{c}{0.126}           \\ \bottomrule
\end{tabular}
}
\caption{Whole-Year Emissions Accounting (Million Tons CO2/MWh)}
\label{tab:noshift_emissions}
\end{table}

\begin{table}[t]
\resizebox{\columnwidth}{!}{%
\begin{tabular}{@{}rcclcclcccccc@{}}
\multicolumn{1}{l}{} & \multicolumn{3}{c}{\textbf{System}}       & \multicolumn{3}{c}{\textbf{DC 103}}       & \multicolumn{2}{c}{\textbf{DC 107}} & \multicolumn{2}{c}{\textbf{DC 204}} & \multicolumn{2}{c}{\textbf{DC 322}} \\ \cmidrule(l){2-13} 
\textbf{}            & Mean          & \multicolumn{2}{c}{SD}    & Mean          & \multicolumn{2}{c}{SD}    & Mean             & SD               & Mean             & SD               & Mean             & SD               \\ \midrule
\textbf{LMCE}        & 0.740         & \multicolumn{2}{c}{0.276} & 0.768         & \multicolumn{2}{c}{0.241} & 0.711            & 0.413            & 0.873            & 0.346            & 0.695            & 0.262            \\
\textbf{ALMCE}       & 0.338         & \multicolumn{2}{c}{0.251} & 0.366         & \multicolumn{2}{c}{0.185} & 0.309            & 0.424            & 0.471            & 0.356            & 0.293            & 0.258            \\
\textbf{ACE}         & 0.342         & \multicolumn{2}{c}{0.170} & 0.342         & \multicolumn{2}{c}{0.170} & 0.342            & 0.170            & 0.342            & 0.170            & 0.342            & 0.170            \\
\textbf{LACE}        & 0.264         & \multicolumn{2}{c}{0.305} & 0.263         & \multicolumn{2}{c}{0.184} & 0.387            & 0.198            & 0.525            & 0.228            & 0.058            & 0.067            \\ \bottomrule
\end{tabular}%
}
\caption{Whole-Year Emissions Intensity (Tons CO2/MWh)}
\label{tab:noshift_intensity}
\end{table}

\paragraph{\textbf{Accounted emissions with different metrics}}
We first consider how the choice of carbon intensity metric impacts the total accounted carbon emissions across the year. Table \ref{tab:noshift_emissions} contains the total system emissions as measured by the generator set-points as well as the accounted emissions for the entire system, the total data center load (denoted by DC load) and for each individual data-center (denoted by DC followed by the corresponding load bus index). In looking at the emissions assigned to all loads in the system, we see that ALMCE, ACE, and LACE lead to accounted emissions that equal the generated emissions value of 15.8 million tons of CO2. LMCE, however, lead to much higher accounted emissions at 33.0 million tons. The trend of much higher emissions with LMCE also holds true for the different data center locations. 

This discrepancy happens because the marginal generators in the RTS-GMLC are much more polluting than the average generation source, and demonstrates the challenges of doing carbon accounting with LMCE. %

\paragraph{\textbf{Carbon intensities with different metrics}}
We next examine the assigned emissions intensities for each carbon metric. Table \ref{tab:noshift_intensity} shows summary statistics of emissions intensity across the all nodes in the system, and for each data-center load individually. We show both the mean value across all hours of the year, as well as the standard deviation of the emissions intensity.
First, we analyze inter-metric intensity differences for the system as a whole. On average, LMCE has the highest intensity values across the whole system, with a mean value of 0.740 Tons CO2/MWh. This is dramatically higher than the next-highest, ACE, with a mean intensity of 0.342 Tons CO2/MWh. ACE shows the least whole-system variability out of all the metrics, likely due to the fact that it does not vary spatially within a given time-step. 

Second, we analyze emissions intensities across the four data-centers. %
As in the whole-system statistics, LMCE assigns the highest average intensity values of any metric to all data-centers. However, the relative magnitude of ALMCE, ACE, and LACE varies greatly between location. Out of these three metrics, LACE has the highest mean value for the DC 107 and DC 204, but the lowest for DC 103 and DC 322. Despite this variability, all of the metrics (save ACE, which is location-agnostic) are in agreement on the ranking of average emissions intensity between data-center locations. DC 204 has the highest emissions intensity, followed by DC 103, DC 107, and finally DC 322.

Finally, we consider the variation in mean carbon intensity across the full power system. Figure \ref{fig:RTS_average} shows the mean nodal emissions intensity values for each metric, with the data-center locations circled in red. We again see that LMCE (left) assigns by far the highest emissions, and that ACE (right) has the same values for all nodes. In this system, LACE shows the highest geographical variability in mean intensity values.

\section{Case Study: Shifting}
We next use the ElectricityEmissions.jl package and the example workflow set out in Section \ref{sec:EA.jl} to analyze what happens when the data centers use different carbon intensity metrics to guide spatio-temporal load shifting (i.e. shifting both between locations and across time steps).

\subsection{Dataset}
We make use of the same modified RTS-GMLC test case as described in the accounting-only case study, which provides a year's worth of hourly load and generation data. Whereas the accounting-only case study added fixed 250MW data-center loads, the data-centers are new allowed to increase or decrease their load by up to 20\% relative to the 250MW baseline in order to minimize emissions.

\subsection{Methodology}

We perform load shifting based on LMCE, ALMCE, ACE, and LACE carbon metrics, within each 24-hour period of year, following the workflow described in Section \ref{sec:workflow} (note that the test-case setup is described in the paragraph above.

\textbf{\emph{1. Initial Emissions Calculations}} We first calculate initial "pre-shift" emissions intensity for each metric, as would be determined from a DA market clearing solution. This provides a forecast of the carbon metric for each node, and at each hour of the day.

\textbf{\emph{2. Solve Data Center Load Shifting Problem}} Next, we optimize the shifting of electricity consumption between the data-centers, which are assumed to be controlled by a single entity. We assume that the data center operator can choose where and when certain computing workloads are being processed. This ability to shift workloads translates into an ability to adjust electricity consumption across locations and time steps. Specifically, the data center operator can shift workloads and associated electricity consumption to align with periods and locations with low carbon intensity, thus minimizing carbon emissions.
We assume that the data centers are able to increase or decrease their electricity consumption by up to 20\% relative to their average consumption (which is set to 250MW in our case) at any given time step, but that the total energy consumption across all data centers for one day (a surrogate for the total computational work that is done) has to remain the same.  The specific mathematical formulation of the data center load shifting problem  
can be found in Appendix \ref{app:load_shifting_formulation}, along with some notes on the potential impact of parameter choice and case-specific modeling decisions. 
We note that in our case study, we assume that only the data centers shift their load, though the simulation setup could be extended to handle (independent) load shifting by other loads as well.

\textbf{\emph{3. Post-Shift Emissions Calculation}}
From the solution to the load-shifting optimization problem, we obtain \textit{estimated} accounted emissions for the data-centers. These estimated accounted emissions are calculated based on the new data-center loading values and the old (pre-shift) carbon intensities.
However, the adjustment to the load consumption then affects the actual load profile in the RT energy market.  We subsequently solve the OPF for the RT market, and calculate the \textit{realized} emissions metrics, after the load shifting has occurred.

\subsection{Results}
\begin{table*}[h]
\resizebox{\textwidth}{!}{%
\begin{tabular}{@{}rcccccccc@{}}
\multicolumn{1}{l}{}          & \multicolumn{3}{c}{\textbf{\begin{tabular}[c]{@{}c@{}}Pre-Shift Accounted \\ Emissions\end{tabular}}} & \textbf{\begin{tabular}[c]{@{}c@{}}Post-Shift Accounted \\ Emissions (Estimated)\end{tabular}} & \multicolumn{3}{c}{\textbf{\begin{tabular}[c]{@{}c@{}}Post-Shift Accounted \\ Emissions (Realized)\end{tabular}}} & \textbf{\begin{tabular}[c]{@{}c@{}}Actual Post-Shift\\ Emissions\end{tabular}} \\
\multicolumn{1}{l}{\textbf{}} & System                        & DC Loads                        & Non-DC Loads                        & DC Loads                                                                                       & System                               & DC Loads                            & Non-DC Loads                         & System                                                                         \\ \midrule
\textbf{LMCE}                 & 33.011                        & 6.692                           & 26.320                              & 6.363 (-4.19\%)                                                                                & 32.689 (-0.98\%)                     & 6.632 (-0.90\%)                     & 26.057 (-1.00\%)                     & 15.669 (-1.00\%)                                                               \\
\textbf{ALMCE}                & 15.828                        & 3.161                           & 12.666                              & 2.865 (-9.36\%)                                                                                & 15.833 (+0.03\%)                     & 3.015 (-4.62\%)                     & 12.818 (+1.20\%)                     & 15.833 (+0.03\%)                                                               \\
\textbf{ACE}                  & 15.828                        & 3.008                           & 12.820                              & 2.824 (-6.12\%)                                                                                & 15.880 (+0.33\%)                     & 2.896 (-3.72\%)                     & 12.985 (+1.29\%)                     & 15.880 (+0.33\%)                                                               \\
\textbf{LACE}                 & 15.828                        & 2.707                           & 13.121                              & 2.370 (-12.45\%)                                                                               & 15.825 (-0.02\%)                     & 2.595 (-4.14\%)                     & 13.230 (+0.83\%)                     & 15.825 (-0.02\%)                                                               \\ \bottomrule
\end{tabular}
}
\caption{Shifting Case Study: Emissions Results (Million Tons CO2)}
\label{tab:shifting_results}
\end{table*}

\begin{figure}[h!]
    \centering
    \includegraphics[width=0.75\linewidth]{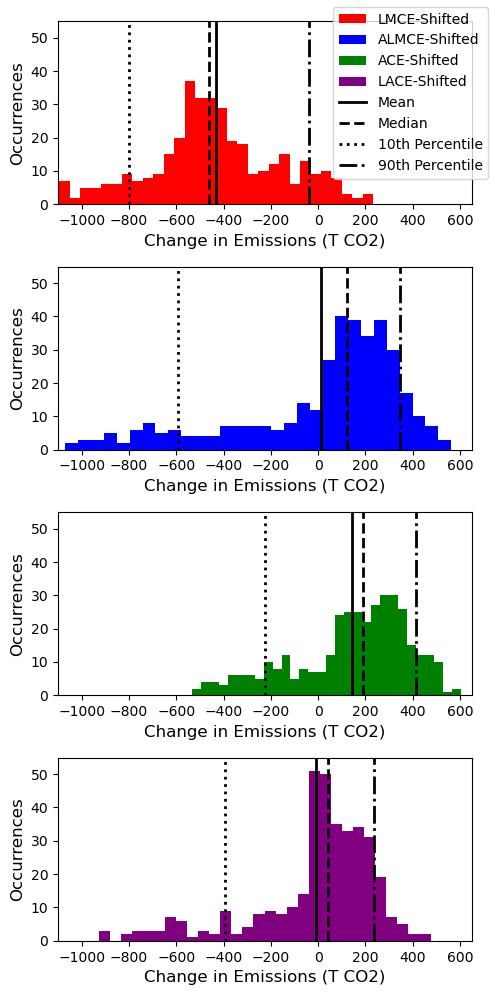}
    \caption{Change in Emissions Resulting From Spatio-Temporal Shifting}
    \label{fig:year_emissions_hists}
\end{figure}

We analyze the emissions outcomes of the load shifting case study to understanding the following for all choices of shifting metric:
\begin{itemize}
    \item The effect of shifting on accounted carbon, as well as on the underlying carbon metric (ie its stability)
    \item The \textit{differential} impact of shifting on carbon responsibility assigned to shifters vs non-shifters
    \item The effect of shifting on ground truth system emissions
    \item The effect of shifting with one metric and accounting with another
\end{itemize}

\paragraph{\textbf{Whole-year results}}
The results of the data center (DC) load-shifting case study are summarized in Table \ref{tab:shifting_results},
and are divided into several categories. 

Starting on the left, we show the pre-shift accounted emissions which are the carbon accounting values prior to any shifting (i.e. they are the same values as in the accounting case study, summarized here for convenience of the reader). 

Next we show the post-shift accounted emissions, which  are divided into "estimated" and "realized". \emph{Estimated} post-shift accounted emissions, which refer to the data-centers' belief about what their accounted emissions will be after shifting, based on their new (post-shift) load values and the old (pre-shift) carbon metric values that they were optimizing for. We only show \emph{estimated} post-shift accounted emissions for the data centers themselves, as these are the only loads that perform load shifting and obtain such estimates. 
The \emph{realized} post-shift emissions are the actual accounted emissions values, determined via post-shift loading and post-shift (i.e. re-calculated) carbon intensity values. 

Finally, we show the total post-shift system emissions, both as an absolute number and a percentage difference to the no-shift total system emissions.

\paragraph{\textbf{Impact of shifting on accounted emissions for data center loads}}
We begin by investigating whether load-shifting leads to a reduction in the accounted emissions assigned to data-center loads.
To analyze this, we first analyze the realized accounted emissions for DC loads. We observe that shifting according to any of the metrics leads to a decrease in the accounted emissions for DC loads, but to varying degrees. The greatest reduction in accounted emissions was achieved by ALMCE, followed by LACE and ACE. The smallest reduction in accounted emissions happened when the data centers shifted according to LMCE. 

Next, comparing the realized accounted emissions for the estimated accounted emissions, we observe a significant discrepancies. Notably, the realized reduction is significantly smaller than the estimated reduction (prior to the final post-shift recalculation of carbon metrics values). Estimated emissions reductions for ALMCE and ACE were ~2x higher than realized, and ~4x and ~3x higher for LMCE and LACE respectively. These discrepancies indicate that the metrics changed due to load shifting, and in ways that reduced the overall effectiveness of the load shifting.

\paragraph{\textbf{Impact of shifting on total system emissions and accounted emissions for non-DC loads}}
Next, we discuss how load shifting with the different carbon intensity metrics changes the overall system emissions, as well as the carbon emissions assigned to non-DC loads.
We first observe that the only metric that significantly reduces the actual post-shift emissions is LMCE. For this metric, we observe a reduction of approximately 1\% in accounted emissions for both non-DC loads and the overall system (similar to the reduction for DC loads). This matches the 1\% reduction in total system emissions, even though the accounted emissions with LMCE are approximately 2x higher than the actual emissions.

In contrast, shifting with ALMCE or LACE barely changes total system emissions, while shifting according to ACE actually increases emissions by 0.33\%. These observations hold true for both the total accounted emissions and the total emissions, as the metrics are designed to ensure that they match. Since the total system emissions hold steady (or increase) for these metrics while fewer emissions are assigned to DC loads, 
the accounted emissions for non-DC loads \textit{increased} by \~1\%. 

In conclusion, these results indicate that even though data centers are able to claim emission reductions after shifting with ALMCE, ACE and LACE, these emission reduction come from reallocating emissions to other loads rather than from actually decreasing total generator emissions. In contrast, when shifting with LMCE, the data centers actually reduced total system emissions.

\paragraph{\textbf{Impact of shifting on individual days}}
While over the course of the whole year, LMCE/LACE-based shifting seemed to decrease emissions and ALMCE/ACE-based shifting seemed to increase them, there is a large amount of variability in the performance of each metric on any particular day. Figure \ref{fig:year_emissions_hists} shows the distribution of the daily changes in system emissions resulting from shifting with the different metrics, as well as the mean, median, 10th percentile and 90th percentile of the reductions. We can clearly see that shifting based on LMCE is much more likely to lead to large carbon emission reductions (i.e. large negative values) and less likely to increase the total carbon emissions from the overall system (i.e. has fewer positive values). We can also clearly see how some of the other metrics have a tendency to increase system carbon emissions rather than reducing them.  %

\begin{figure}%
    \centering
    \includegraphics[width=\linewidth]{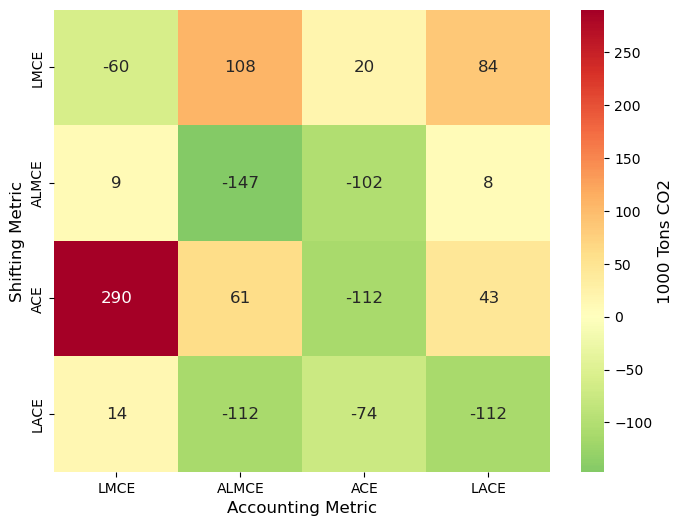}
    \caption{Cross-Metric Shifting/Accounting: Change in Accounted Data-Center Emissions}
    \label{fig:cross_metric_heatmap}
\end{figure}

\paragraph{\textbf{Impact of shifting with one metric and accounting with another}}
Finally, we briefly examine the outcome of shifting based on one metric and accounting using another (e.g. determining data-center load shifting with LMCE, then assigning emissions using ACE).  Figure \ref{fig:cross_metric_heatmap} shows the absolute change in assigned data-center emissions, defined as the difference between pre- and post-shift accounted emissions (measured by the accounting metric). 

As is to be expected, when the same metric is used for shifting and accounting, we see the greatest reduction in accounted emissions. There is great variation in the 'cross-metric' behavior of different "shifting metric"-"accounting metric" pairs. For many of them, we see that shifting with one metric (e.g. LMCE) increases the accounted carbon emissions when a different metric is used for accounting. This indicates that companies or organizations who want to not only reduce overall carbon emissions, but also the carbon emissions that are allocated to them, have strong incentives to shift according to the same metric that is used for accounting.  %

\section{Conclusions}
Currently, there exists multiple competing definitions of the carbon intensity of electricity. In this paper, we seek to analyze benefits and drawbacks of existing, emerging  and newly proposed carbon intensity metrics for two important use cases, namely load shifting and carbon accounting.
To achieve this goal, we provide an overview and qualitative comparison of different metrics, and implement them in a  new software package, ElectricityEmissions.jl. This software package provides tools for the calculation and analysis of several carbon emissions intensity metrics. %

Leveraging this package, we compare the two most established carbon intensity metrics, namely average carbon emissions (ACE) and locational marginal carbon emissions (LMCE), against the recently emerging locational average carbon emissions (LACE) and a new metric proposed in this paper, which we refer to as 
adjusted locational marginal carbon emissions (ALMCE). We analyze the outcomes of using those metrics both in the context of accounting and load shifting.

In our case study, we find that load shifting based on the ACE, ALMCE and LACE intensity metrics has very little impact on overall carbon emission and instead tends to shift emissions from the data centers (which actively seek to minimize their emissions according to these metrics) to non-data center loads. In contrast, shifting according to LMCE reduces emissions for both data centers, non-data center loads, and the overall system. However, it is challenging to do accounting with LMCE, as the allocated carbon emissions are more than two times higher than the actual carbon emissions of the system.

The above findings are specific to the case we study, and further work is required to assess the effectiveness of the different metrics across a larger range of case studies. However, our results demonstrate that load shifting is not always a productive method to reduce short-term carbon emissions, and can be counter-productive if the carbon intensity metric is not chosen carefully.

In future work, we plan to expand the package to include other metrics and other functionality, allowing input from the community. In particular, we want to add support quadratic generator cost functions and alternative OPF formulations. We also wish to design new metrics with nice properties for both accounting and load shifting, and perform more thorough analysis of the differences between metrics on more realistic power system test cases and more realistic load-shifting setups.

\appendix
\section{Appendix}

\subsection{DCOPF Formulation}
\label{app:dcopf_formulation}
We use the following DCOPF formulation with piecewise linear generator cost to dispatch generation in ElectricityEmissions.jl, and for the linear generation sensitivity calculations necessary for the LMCE and ALMCE carbon metrics:

\begin{subequations}
\label{Eq:DCOPF}
\begin{align}
    & \min_{\theta, p_G, C_G} && \sum_j{C_{G,j}} \\
    & \text{s.t.} && C_{G,j} \geq a_{j,k}p_{G,j} + b_{j,k} & \hspace{-15em} \forall j=1...N_G,k=1...N_{ct} \label{eq:dcopf_pwl} \\
    &&& \sum_{l \in G_i}p_{G,l}  - \!\!\! \sum_{l \in D_i}p_{D,l} = \!\!\!\!\!\! \sum_{j:(i,j)\in L} \!\!\!\!\!\! - \beta_{ij}(\theta_i \!- \!\theta_j) &  \hspace{-0.5em} \forall i=1...N \label{eq:dcopf_conservation} \\
    &&& -P_{ij}^{lim} \leq -\beta_{ij}(\theta_i - \theta_j) \leq P_{ij}^{lim} &  \hspace{-2em} \forall (i,j)\in L \label{eq:dcopf_flowlim}\\
    &&& P_{G,j}^{min} \leq p_{G,j} \leq P_{G,j}^{max} & \hspace{-2em} \forall j=1...N_g \label{eq:dcopf_genlim}\\
    &&& \theta_{ref} = 0 \label{eq:dcopf_refbus}
\end{align}
\end{subequations}
Here, the decision variables $\theta$, $p_G$, and $C_G$ refer to nodal voltage angles, generator set-points, and generator costs respectively. There are $N$ buses in the system and $N_G$ generators. The system is connected by lines $(i,j)\in L$, specifying the start and end buses $i$ and $j$. 

Each generator $j$ has a piecewise-linear cost function with $N_{ct}$ terms. Equation \eqref{eq:dcopf_pwl} constrains the cost of each generator $j$ to be above all of the $k$ linear functions that define its segments, with $a_{j,k}$ and $b_{jk}$ representing the slope and intercept of the function. Equation \eqref{eq:dcopf_conservation} is the standard nodal power conservation constraint, referencing the subset of generators $G_i\subset\mathcal{G}$, loads $D_i\subset\mathcal{D}$, and lines $(i,j)\in L$ connected at bus $i$. Equation \eqref{eq:dcopf_flowlim} limits the flow on each line $(i,j)$ to be below the maximum capacity $P_{i,j}^{lim}$, and Equation \eqref{eq:dcopf_genlim} sets limits for each generator $j$ to be between $P_{G,j}^{min}$ and $P_{G,j}^{max}$. Finally, Equation \eqref{eq:dcopf_refbus} sets the voltage angle of the reference bus to zero.

\subsection{LMCE Calculation}
\label{app:LMCE_calc}
LMCE is calculated using linear sensitivity analysis for the DCOPF solution, and describes the effect on generation and emissions due to a small change in load. For ElectricityEmissions.jl, we use the general method set forth in \cite{lindberg2021guide} to calculate this sensitivity. However, we make several small modifications in order to deal with the new constraints and variables introduced by the piecewise-linear cost functions used in our DCOPF formulation (where \cite{lindberg2021guide} simply assumed linear costs).

The DCOPF formulation, as described in Appendix \ref{app:dcopf_formulation}, has $n = N + 2N_G$ decision variables, where $N$ is the number of buses and $N_G$ is the number of generators. For each bus, the voltage angle $\Theta_i$ must be determined, and for each generator $j$, the active power generation $p_{G,j}$ and cost $C_{G,j}$ must be determined. 

For any instance of the DCOPF problem, which is a linear program, there exists at least on \emph{basic optimal solution}. At this solution, we can identify $n$ linearly independent constraints that are being satisfied with equality.
For this solution, the following linear system can be formed
\begin{equation}
Ax^*=b  
\label{lin_system}
\end{equation}
where $A\in \mathbb{R}^{n \times n}$ contains the problem's equality and binding inequality constraints, $x^* \in \mathbb{R}^n$ contains the (unique) decision variable values at optimality, and $b\in \mathbb{R}^n$ contains the standard-form RHS constraint bounds.

We note that not all solutions returned by a solver may be basic feasible, and it is possible that the problem exhibits degeneracy.
One notable potential source of degeneracy is a non-unique solution due to the presence of more than one generator with the same cost function at a given node, leading to an A matrix that is not full rank. As noted in the first step of the example \textit{ElectricityEmissions.jl} workflow (Section \ref{sec:workflow}), this is easily remedied by adding a small amount of noise to such duplicate cost functions.

As we are interested in the relationship between changes in load and changes in generation dispatch, we instead make use of the system
\begin{equation}
A\Delta x = \Delta b
\label{lin_delta_system}
\end{equation}
which is implied by Eq. \ref{lin_system}. The constraints and variables are input to the system such that
\begin{equation}
    A \begin{bmatrix}\Delta \Theta \\ \Delta p_G \\ \Delta C_G\end{bmatrix} = \begin{bmatrix} \Delta p_D \\ 0\end{bmatrix}
\end{equation}
where $\Theta \in \mathbb{R}^{N}$ refers to the vector of bus voltage angles, $p_G \in \mathbb{R}^{N_G} $ the vector of generator set-points, and $C_g \in \mathbb{R}^{N_G}$ the generator cost values. $p_D \in \mathbb{R}^{N} $ is the vector containing total load at each bus. 
In this formulation, the nodal power balance constraints, which are the only constraints to include $P_D$ variables, make up the first $N$ rows of the $A$ matrix. For ease in calculation, they are ordered according to their bus index. The final $n-N$ rows of $A$ correspond to the slack bus equality constraints, and all binding inequality constraints for which the order does not matter. This system of equations is identical to that used in \cite{lindberg2021guide}, save for the addition of the $C_G$ variables representing the piecewise-linear generator costs, and any binding constraints on these variables (encoded in the $A$ matrix).

Following the formation of this system, we then invert the $A$ matrix, and solve for $\Delta p_D$. For compactness, we define the $B$ matrix to be the first $N$ columns and rows $N+1$ through $N+N_g$ of $A^{-1}$ These are simply the columns of $A^{-1}$ aligning with $\Delta p_D$ and the rows aligning with $\Delta p_G$ . We can then write
\begin{equation}
    \Delta p_G = B \Delta p_D.
    \label{eq:B_eq}
\end{equation}
To determine the change in generation $\Delta p_{G,j}$ at generator $j$ resulting from one unit of additional load at bus $i$, we set %
the $i$th entry of $\Delta p_D$ to 1 and all others to 0.
We can then perform the final LMCE calculation as
\begin{equation}
    LMCE_i = \sum_{j\in\mathcal{G}} e_{G,j} \Delta p_{G,j},
\end{equation}
where $e_{G,j}$ generator emissions intensity of generator $j$.

\subsection{Load Shifting Formulation}
\label{app:load_shifting_formulation}

We perform spatio-temporal shifting, in which a group of data-centers are allowed to shift their load both to other data-center locations and other times of day. The spatio-temporal shifting problem is formulated as follows:
\begin{align}
    & min_{d_{i,t}} &&  \sum_{i,t}{e_{D,i,t}p_{D,i,t}} \label{eq:obj}\\
    & s.t.      &&  \sum_{i,t}{p_{D,i,t}} = N_d \cdot N_t \cdot D_{nom} \label{eq:dc_STshift_bal} \\
    &&& (1-\epsilon)\cdot D_{nom} \leq d_{i,t} \leq (1+\epsilon)\cdot D_{nom}, \nonumber\\ &&&\qquad\qquad\qquad\qquad\qquad\forall i=1....N_d, t=1....N_t,\label{eq:dc_STshift_lim} 
\end{align}
Here, $p_{D,i,t}$ and $e_{D,i,t}$ represent the load and carbon intensity at data-center $i$ and hour $t$ of the day. $N_d$ and $N_t$ are the number of data-centers and number of time-steps, in this case set to 4 and 24 respectively. $D_{nom}$ is the nominal data-center load (here set to 250MW), and $\epsilon$ is the data-center load flexibility (here set to 0.2). Eq. \eqref{eq:dc_STshift_bal} enforces that the total data-center energy consumption (across all locations and time-steps) is the same as if all data-centers were at nominal load for the entire period. Eq. \eqref{eq:dc_STshift_lim} defines a feasible range within which data center load can be adjusted up or down at a particular time-step (where the size of this range is controlled by $\epsilon$) . For our choice of parameters (4 data-centers with a constant nominal load of 250MW over 24 hours), the data centers have a total daily energy consumption of 24,000MWh.

We note that shifting outcomes, including those presented in this work, are likely to vary significantly depending on the specific implementation. Of particular importance are factors that influence the size of load-shifts, such as constraints on data-center flexibility ($\epsilon$ in our model), as well as case-specific shifting costs (e.g. from communications overhead) which we do not consider here. As a result of such effects, we do not include the shifting formulation employed in this work as part of the \textit{ElectricityEmissions.jl} package, under the assumption that users may wish to model a variety of flexible load types with specific qualities/requirements. 

\vspace{+5pt}
\noindent Los Alamos Unlimited Release LA-UR-24-31644. Reviewed for release outside the Laboratory with no distribution restrictions.

\begin{acks}
This work is funded through the National Science Foundation, under awards \#2328160 and \#2325956.
\end{acks}
\bibliographystyle{ACM-Reference-Format}
\bibliography{references}


\begin{thebibliography}{39}


\ifx \showCODEN    \undefined \def \showCODEN     #1{\unskip}     \fi
\ifx \showDOI      \undefined \def \showDOI       #1{#1}\fi
\ifx \showISBNx    \undefined \def \showISBNx     #1{\unskip}     \fi
\ifx \showISBNxiii \undefined \def \showISBNxiii  #1{\unskip}     \fi
\ifx \showISSN     \undefined \def \showISSN      #1{\unskip}     \fi
\ifx \showLCCN     \undefined \def \showLCCN      #1{\unskip}     \fi
\ifx \shownote     \undefined \def \shownote      #1{#1}          \fi
\ifx \showarticletitle \undefined \def \showarticletitle #1{#1}   \fi
\ifx \showURL      \undefined \def \showURL       {\relax}        \fi
\providecommand\bibfield[2]{#2}
\providecommand\bibinfo[2]{#2}
\providecommand\natexlab[1]{#1}
\providecommand\showeprint[2][]{arXiv:#2}

\bibitem[Barrows et~al\mbox{.}(2020)]%
        {RTSGMLC}
\bibfield{author}{\bibinfo{person}{Clayton Barrows}, \bibinfo{person}{Aaron Bloom}, \bibinfo{person}{Ali Ehlen}, \bibinfo{person}{Jussi Ikäheimo}, \bibinfo{person}{Jennie Jorgenson}, \bibinfo{person}{Dheepak Krishnamurthy}, \bibinfo{person}{Jessica Lau}, \bibinfo{person}{Brendan McBennett}, \bibinfo{person}{Matthew O’Connell}, \bibinfo{person}{Eugene Preston}, \bibinfo{person}{Andrea Staid}, \bibinfo{person}{Gord Stephen}, {and} \bibinfo{person}{Jean-Paul Watson}.} \bibinfo{year}{2020}\natexlab{}.
\newblock \showarticletitle{The IEEE Reliability Test System: A Proposed 2019 Update}.
\newblock \bibinfo{journal}{\emph{IEEE Trans. on Power Systems}} \bibinfo{volume}{35}, \bibinfo{number}{1} (\bibinfo{year}{2020}), \bibinfo{pages}{119--127}.
\newblock
\urldef\tempurl%
\url{https://doi.org/10.1109/TPWRS.2019.2925557}
\showDOI{\tempurl}


\bibitem[Bialek(1996)]%
        {bialek1996tracing}
\bibfield{author}{\bibinfo{person}{Janusz Bialek}.} \bibinfo{year}{1996}\natexlab{}.
\newblock \showarticletitle{Tracing the flow of electricity}.
\newblock \bibinfo{journal}{\emph{IEE Proceedings-Generation, Transmission and Distribution}} \bibinfo{volume}{143}, \bibinfo{number}{4} (\bibinfo{year}{1996}), \bibinfo{pages}{313--320}.
\newblock


\bibitem[Callaway et~al\mbox{.}(2018)]%
        {callaway2018location}
\bibfield{author}{\bibinfo{person}{Duncan~S Callaway}, \bibinfo{person}{Meredith Fowlie}, {and} \bibinfo{person}{Gavin McCormick}.} \bibinfo{year}{2018}\natexlab{}.
\newblock \showarticletitle{Location, location, location: The variable value of renewable energy and demand-side efficiency resources}.
\newblock \bibinfo{journal}{\emph{Journal of the Association of Environmental and Resource Economists}} \bibinfo{volume}{5}, \bibinfo{number}{1} (\bibinfo{year}{2018}), \bibinfo{pages}{39--75}.
\newblock


\bibitem[Carleton~Coffrin(2024)]%
        {pglib_opf}
\bibfield{author}{\bibinfo{person}{Ray~Zimmerman Carleton~Coffrin}.} \bibinfo{year}{2024}\natexlab{}.
\newblock \bibinfo{title}{Benchmarks for the Optimal Power Flow Problem}.
\newblock
\newblock
\urldef\tempurl%
\url{https://github.com/power-grid-lib/pglib-opf}
\showURL{%
\tempurl}


\bibitem[Chen et~al\mbox{.}(2023a)]%
        {chen2023towards}
\bibfield{author}{\bibinfo{person}{Xin Chen}, \bibinfo{person}{Hungpo Chao}, \bibinfo{person}{Wenbo Shi}, {and} \bibinfo{person}{Na Li}.} \bibinfo{year}{2023}\natexlab{a}.
\newblock \showarticletitle{Towards carbon-free electricity: A comprehensive flow-based framework for power grid carbon accounting and decarbonization}.
\newblock \bibinfo{journal}{\emph{arXiv preprint arXiv:2308.03268}} (\bibinfo{year}{2023}).
\newblock


\bibitem[Chen et~al\mbox{.}(2023c)]%
        {chen2023carbon}
\bibfield{author}{\bibinfo{person}{Xin Chen}, \bibinfo{person}{Andy Sun}, \bibinfo{person}{Wenbo Shi}, {and} \bibinfo{person}{Na Li}.} \bibinfo{year}{2023}\natexlab{c}.
\newblock \showarticletitle{Carbon-Aware Optimal Power Flow}.
\newblock \bibinfo{journal}{\emph{arXiv preprint arXiv:2308.03240}} (\bibinfo{year}{2023}).
\newblock


\bibitem[Chen et~al\mbox{.}(2023b)]%
        {deka2023contributions}
\bibfield{author}{\bibinfo{person}{Yize Chen}, \bibinfo{person}{Deepyjoti Deka}, {and} \bibinfo{person}{Yuanyuan Shi}.} \bibinfo{year}{2023}\natexlab{b}.
\newblock \showarticletitle{Contributions of Individual Generators to Nodal Carbon Emissions}.
\newblock \bibinfo{journal}{\emph{arXiv preprint arXiv:2311.03712}} (\bibinfo{year}{2023}).
\newblock


\bibitem[Chen and Dhople(2019)]%
        {dhople2019tracing}
\bibfield{author}{\bibinfo{person}{Yu~Christine Chen} {and} \bibinfo{person}{Sairaj~V Dhople}.} \bibinfo{year}{2019}\natexlab{}.
\newblock \showarticletitle{Tracing power with circuit theory}.
\newblock \bibinfo{journal}{\emph{IEEE Transactions on Smart Grid}} \bibinfo{volume}{11}, \bibinfo{number}{1} (\bibinfo{year}{2019}), \bibinfo{pages}{138--147}.
\newblock


\bibitem[Coffrin et~al\mbox{.}(2018)]%
        {powermodels_jl}
\bibfield{author}{\bibinfo{person}{Carleton Coffrin}, \bibinfo{person}{Russell Bent}, \bibinfo{person}{Kaarthik Sundar}, \bibinfo{person}{Yeesian Ng}, {and} \bibinfo{person}{Miles Lubin}.} \bibinfo{year}{2018}\natexlab{}.
\newblock \showarticletitle{PowerModels.jl: An Open-Source Framework for Exploring Power Flow Formulations}. In \bibinfo{booktitle}{\emph{2018 Power Systems Computation Conference (PSCC)}}. \bibinfo{pages}{1--8}.
\newblock
\urldef\tempurl%
\url{https://doi.org/10.23919/PSCC.2018.8442948}
\showDOI{\tempurl}


\bibitem[Corradi(2023)]%
        {electricitymapsblog2023marginal}
\bibfield{author}{\bibinfo{person}{Olivier Corradi}.} \bibinfo{year}{2023}\natexlab{}.
\newblock \bibinfo{title}{Marginal vs Average: Which One to Use for Real-time Decisions?}
\newblock
\newblock
\urldef\tempurl%
\url{https://www.electricitymaps.com/blog/marginal-vs-average-realtime-decision-making}
\showURL{%
\tempurl}


\bibitem[{Electricitymaps}({[n.\,d.]})]%
        {electricitymaps}
\bibfield{author}{\bibinfo{person}{{Electricitymaps}}.} \bibinfo{year}{[n.\,d.]}\natexlab{}.
\newblock \bibinfo{title}{The leading API for granular electricity data Reduce carbon emissions with actionable electricity data}.
\newblock \bibinfo{howpublished}{\url{https://www.electricitymaps.com}}.
\newblock


\bibitem[{GHG Protocol}(2015)]%
        {ghg2015}
\bibfield{author}{\bibinfo{person}{{GHG Protocol}}.} \bibinfo{year}{2015}\natexlab{}.
\newblock \bibinfo{booktitle}{\emph{{Greenhouse Gas Protocol Scope 2 Guidance. An amendment to the GHG Protocol Corporate Standard}}}.
\newblock \bibinfo{type}{{T}echnical {R}eport}.
\newblock


\bibitem[Harris et~al\mbox{.}(2015)]%
        {Harris2015Residential}
\bibfield{author}{\bibinfo{person}{A. Harris}, \bibinfo{person}{M. Rogers}, \bibinfo{person}{Carol~J. Miller}, \bibinfo{person}{S. McElmurry}, {and} \bibinfo{person}{Caisheng Wang}.} \bibinfo{year}{2015}\natexlab{}.
\newblock \showarticletitle{Residential emissions reductions through variable timing of electricity consumption}.
\newblock \bibinfo{journal}{\emph{Applied Energy}}  \bibinfo{volume}{158} (\bibinfo{year}{2015}), \bibinfo{pages}{484--489}.
\newblock
\urldef\tempurl%
\url{https://doi.org/10.1016/J.APENERGY.2015.08.042}
\showDOI{\tempurl}


\bibitem[Hawkes(2014)]%
        {hawkes2014long}
\bibfield{author}{\bibinfo{person}{Adam~D Hawkes}.} \bibinfo{year}{2014}\natexlab{}.
\newblock \showarticletitle{Long-run marginal CO2 emissions factors in national electricity systems}.
\newblock \bibinfo{journal}{\emph{Applied Energy}}  \bibinfo{volume}{125} (\bibinfo{year}{2014}), \bibinfo{pages}{197--205}.
\newblock


\bibitem[Inc.(2023)]%
        {apple_clean_energy_charging}
\bibfield{author}{\bibinfo{person}{Apple Inc.}} \bibinfo{year}{2023}\natexlab{}.
\newblock \bibinfo{title}{Use Clean Energy Charging on your iPhone}.
\newblock
\newblock
\urldef\tempurl%
\url{https://support.apple.com/en-us/108068}
\showURL{%
\tempurl}


\bibitem[{Internal Revenue Service}(2023)]%
        {proposed-rule}
\bibfield{author}{\bibinfo{person}{{Internal Revenue Service}}.} \bibinfo{year}{2023}\natexlab{}.
\newblock \bibinfo{title}{{Proposed Rule: Section 45V Credit for Production of Clean Hydrogen; Section 48(a)(15) Election To Treat Clean Hydrogen Production Facilities as Energy Property}}.
\newblock \bibinfo{howpublished}{\url{https://www.federalregister.gov/documents/2023/12/26/2023-28359/section-45v-credit-for-production-of-clean-hydrogen-section-48a15- \ election-to-treat-clean-hydrogen}}.
\newblock


\bibitem[Kang et~al\mbox{.}(2015)]%
        {kang2015carbon}
\bibfield{author}{\bibinfo{person}{Chongqing Kang}, \bibinfo{person}{Tianrui Zhou}, \bibinfo{person}{Qixin Chen}, \bibinfo{person}{Jianhui Wang}, \bibinfo{person}{Yanlong Sun}, \bibinfo{person}{Qing Xia}, {and} \bibinfo{person}{Huaguang Yan}.} \bibinfo{year}{2015}\natexlab{}.
\newblock \showarticletitle{Carbon emission flow from generation to demand: A network-based model}.
\newblock \bibinfo{journal}{\emph{IEEE Transactions on Smart Grid}} \bibinfo{volume}{6}, \bibinfo{number}{5} (\bibinfo{year}{2015}), \bibinfo{pages}{2386--2394}.
\newblock


\bibitem[Kirschen and Strbac(1999)]%
        {kirschen1999tracing}
\bibfield{author}{\bibinfo{person}{Daniel Kirschen} {and} \bibinfo{person}{Goran Strbac}.} \bibinfo{year}{1999}\natexlab{}.
\newblock \showarticletitle{Tracing active and reactive power between generators and loads using real and imaginary currents}.
\newblock \bibinfo{journal}{\emph{IEEE Transactions on Power Systems}} \bibinfo{volume}{14}, \bibinfo{number}{4} (\bibinfo{year}{1999}), \bibinfo{pages}{1312--1319}.
\newblock


\bibitem[Lin et~al\mbox{.}(2012)]%
        {Lin2012}
\bibfield{author}{\bibinfo{person}{Minghong Lin}, \bibinfo{person}{Zhenhua Liu}, \bibinfo{person}{Adam Wierman}, {and} \bibinfo{person}{Lachlan L.~H. Andrew}.} \bibinfo{year}{2012}\natexlab{}.
\newblock \showarticletitle{Online algorithms for geographical load balancing}. In \bibinfo{booktitle}{\emph{2012 International Green Computing Conference (IGCC)}}. \bibinfo{pages}{1--10}.
\newblock
\urldef\tempurl%
\url{https://doi.org/10.1109/IGCC.2012.6322266}
\showDOI{\tempurl}


\bibitem[Lindberg et~al\mbox{.}(2021a)]%
        {lindberg2021guide}
\bibfield{author}{\bibinfo{person}{Julia Lindberg}, \bibinfo{person}{Yasmine Abdennadher}, \bibinfo{person}{Jiaqi Chen}, \bibinfo{person}{Bernard~C Lesieutre}, {and} \bibinfo{person}{Line Roald}.} \bibinfo{year}{2021}\natexlab{a}.
\newblock \showarticletitle{A guide to reducing carbon emissions through data center geographical load shifting}. In \bibinfo{booktitle}{\emph{Proceedings of the Twelfth ACM International Conference on Future Energy Systems}}. \bibinfo{pages}{430--436}.
\newblock


\bibitem[Lindberg et~al\mbox{.}(2021b)]%
        {lindberg2020environmental}
\bibfield{author}{\bibinfo{person}{Julia Lindberg}, \bibinfo{person}{Bernard~C Lesieutre}, {and} \bibinfo{person}{Line Roald}.} \bibinfo{year}{2021}\natexlab{b}.
\newblock \showarticletitle{The Environmental Potential of Hyper-Scale Data Centers: Using Locational Marginal CO $ \_2 $ Emissions to Guide Geographical Load Shifting}.
\newblock \bibinfo{journal}{\emph{{Hawaii International Conference on System Sciences}}} (\bibinfo{year}{2021}).
\newblock


\bibitem[Lindberg et~al\mbox{.}(2022)]%
        {lindberg2022using}
\bibfield{author}{\bibinfo{person}{Julia Lindberg}, \bibinfo{person}{Bernard~C Lesieutre}, {and} \bibinfo{person}{Line~A Roald}.} \bibinfo{year}{2022}\natexlab{}.
\newblock \showarticletitle{Using geographic load shifting to reduce carbon emissions}.
\newblock \bibinfo{journal}{\emph{Electric Power Systems Research}}  \bibinfo{volume}{212} (\bibinfo{year}{2022}), \bibinfo{pages}{108586}.
\newblock


\bibitem[Maji et~al\mbox{.}(2023)]%
        {Maji2023Bringing}
\bibfield{author}{\bibinfo{person}{Diptyaroop Maji}, \bibinfo{person}{Benedikt Pfaff}, \bibinfo{person}{Vipin~P R}, \bibinfo{person}{Rajagopal Sreenivasan}, \bibinfo{person}{V. Firoiu}, \bibinfo{person}{Sreeram Iyer}, \bibinfo{person}{Colleen Josephson}, \bibinfo{person}{Zhelong Pan}, {and} \bibinfo{person}{R. Sitaraman}.} \bibinfo{year}{2023}\natexlab{}.
\newblock \showarticletitle{Bringing Carbon Awareness to Multi-cloud Application Delivery}.
\newblock \bibinfo{journal}{\emph{Proceedings of the 2nd Workshop on Sustainable Computer Systems}} (\bibinfo{year}{2023}).
\newblock
\urldef\tempurl%
\url{https://doi.org/10.1145/3604930.3605711}
\showDOI{\tempurl}


\bibitem[Mata et~al\mbox{.}(2020)]%
        {Mata2020A}
\bibfield{author}{\bibinfo{person}{É. Mata}, \bibinfo{person}{Jonas Ottosson}, {and} \bibinfo{person}{J. Nilsson}.} \bibinfo{year}{2020}\natexlab{}.
\newblock \showarticletitle{A review of flexibility of residential electricity demand as climate solution in four EU countries}.
\newblock \bibinfo{journal}{\emph{Environmental Research Letters}}  \bibinfo{volume}{15} (\bibinfo{year}{2020}).
\newblock
\urldef\tempurl%
\url{https://doi.org/10.1088/1748-9326/ab7950}
\showDOI{\tempurl}


\bibitem[{National Energy System Operator}({[n.\,d.]})]%
        {carbonintensity}
\bibfield{author}{\bibinfo{person}{{National Energy System Operator}}.} \bibinfo{year}{[n.\,d.]}\natexlab{}.
\newblock \bibinfo{title}{Carbon Intensity API}.
\newblock \bibinfo{howpublished}{\url{https://carbonintensity.org.uk/}}.
\newblock


\bibitem[{National Grid}({[n.\,d.]})]%
        {greenlight}
\bibfield{author}{\bibinfo{person}{{National Grid}}.} \bibinfo{year}{[n.\,d.]}\natexlab{}.
\newblock \bibinfo{title}{The Green Light Signal}.
\newblock \bibinfo{howpublished}{\url{https://www.nationalgrid.com/greenlightsignal}}.
\newblock


\bibitem[{Office of Energy Policy} and {Systems Analysis, U.S. Department of Energy}(2016)]%
        {environmentbaseline}
\bibfield{author}{\bibinfo{person}{{Office of Energy Policy}} {and} \bibinfo{person}{{Systems Analysis, U.S. Department of Energy}}.} \bibinfo{year}{2016}\natexlab{}.
\newblock \showarticletitle{{Environment Baseline, Volume 1: Greenhouse Gas Emissions from the U.S. Power Sector}}.
\newblock  (\bibinfo{year}{2016}).
\newblock


\bibitem[{PJM}({[n.\,d.]})]%
        {pjm}
\bibfield{author}{\bibinfo{person}{{PJM}}.} \bibinfo{year}{[n.\,d.]}\natexlab{}.
\newblock \bibinfo{title}{{Five Minute Marginal Emission Rates}}.
\newblock \bibinfo{howpublished}{\url{https://dataminer2.pjm.com/feed/fivemin_marginal_emissions/definition}}.
\newblock


\bibitem[PJM(2024)]%
        {pjm_lmce}
\bibfield{author}{\bibinfo{person}{PJM}.} \bibinfo{year}{2024}\natexlab{}.
\newblock \bibinfo{title}{5 Minute Marginal Emissions Rates}.
\newblock
\newblock
\urldef\tempurl%
\url{https://dataminer2.pjm.com/feed/fivemin_marginal_emissions}
\showURL{%
\tempurl}


\bibitem[Radovanovic et~al\mbox{.}(2022)]%
        {Radovanovic2021}
\bibfield{author}{\bibinfo{person}{Ana Radovanovic}, \bibinfo{person}{Ross Koningstein}, \bibinfo{person}{Ian Schneider}, \bibinfo{person}{Bokan Chen}, \bibinfo{person}{Alexandre Duarte}, \bibinfo{person}{Binz Roy}, \bibinfo{person}{Diyue Xiao}, \bibinfo{person}{Maya Haridasan}, \bibinfo{person}{Patrick Hung}, \bibinfo{person}{Nick Care}, \bibinfo{person}{Saurav Talukdar}, \bibinfo{person}{Eric Mullen}, \bibinfo{person}{Kendal Smith}, \bibinfo{person}{Mariellen Cottman}, {and} \bibinfo{person}{Walfredo Cirne}.} \bibinfo{year}{2022}\natexlab{}.
\newblock \showarticletitle{Carbon-Aware Computing for Datacenters}.
\newblock \bibinfo{journal}{\emph{IEEE Transactions on Power Systems}} (\bibinfo{year}{2022}), \bibinfo{pages}{1--1}.
\newblock
\urldef\tempurl%
\url{https://doi.org/10.1109/TPWRS.2022.3173250}
\showDOI{\tempurl}


\bibitem[Richardson(2023)]%
        {wattimeblog2023marginal}
\bibfield{author}{\bibinfo{person}{Henry Richardson}.} \bibinfo{year}{2023}\natexlab{}.
\newblock \bibinfo{title}{Is Your Goal Real-world Impact? Then Use Marginal Emissions}.
\newblock
\newblock
\urldef\tempurl%
\url{https://www.watttime.org/news/is-your-goal-real-world-impactthen-use-marginal-emissions/}
\showURL{%
\tempurl}


\bibitem[Ricks et~al\mbox{.}(2024)]%
        {RICKS2024114119}
\bibfield{author}{\bibinfo{person}{Wilson Ricks}, \bibinfo{person}{Pieter Gagnon}, {and} \bibinfo{person}{Jesse~D. Jenkins}.} \bibinfo{year}{2024}\natexlab{}.
\newblock \showarticletitle{Short-run marginal emission factors neglect impactful phenomena and are unsuitable for assessing the power sector emissions impacts of hydrogen electrolysis}.
\newblock \bibinfo{journal}{\emph{Energy Policy}}  \bibinfo{volume}{189} (\bibinfo{year}{2024}), \bibinfo{pages}{114119}.
\newblock
\showISSN{0301-4215}
\urldef\tempurl%
\url{https://doi.org/10.1016/j.enpol.2024.114119}
\showDOI{\tempurl}


\bibitem[Ricks et~al\mbox{.}(2022)]%
        {Ricks2022Minimizing}
\bibfield{author}{\bibinfo{person}{W. Ricks}, \bibinfo{person}{Qingyu Xu}, {and} \bibinfo{person}{J. Jenkins}.} \bibinfo{year}{2022}\natexlab{}.
\newblock \showarticletitle{Minimizing emissions from grid-based hydrogen production in the United States}.
\newblock \bibinfo{journal}{\emph{Environmental Research Letters}}  \bibinfo{volume}{18} (\bibinfo{year}{2022}).
\newblock
\urldef\tempurl%
\url{https://doi.org/10.1088/1748-9326/acacb5}
\showDOI{\tempurl}


\bibitem[Sukprasert et~al\mbox{.}(2024)]%
        {sukprasert2024implications}
\bibfield{author}{\bibinfo{person}{Thanathorn Sukprasert}, \bibinfo{person}{Noman Bashir}, \bibinfo{person}{Abel Souza}, \bibinfo{person}{David Irwin}, {and} \bibinfo{person}{Prashant Shenoy}.} \bibinfo{year}{2024}\natexlab{}.
\newblock \showarticletitle{On the Implications of Choosing Average versus Marginal Carbon Intensity Signals on Carbon-aware Optimizations}. In \bibinfo{booktitle}{\emph{Proceedings of the 15th ACM International Conference on Future and Sustainable Energy Systems}}. \bibinfo{pages}{422--427}.
\newblock


\bibitem[{WattTime}({[n.\,d.]})]%
        {wattime}
\bibfield{author}{\bibinfo{person}{{WattTime}}.} \bibinfo{year}{[n.\,d.]}\natexlab{}.
\newblock \bibinfo{title}{We're a nonprofit raising awareness about these solutions and providing data \& technical assistance to anyone trying to implement them.}
\newblock \bibinfo{howpublished}{\url{https://watttime.org/}}.
\newblock


\bibitem[{Wisconsin Power Optimization Lab (WISPO-POP)}({[n.\,d.]})]%
        {powerplots}
\bibfield{author}{\bibinfo{person}{{Wisconsin Power Optimization Lab (WISPO-POP)}}.} \bibinfo{year}{[n.\,d.]}\natexlab{}.
\newblock \bibinfo{title}{{PowerPlots.jl}}.
\newblock \bibinfo{howpublished}{\url{https://github.com/WISPO-POP/PowerPlots.jl}}.
\newblock


\bibitem[Zhang et~al\mbox{.}(2020)]%
        {Zhang2020Flexible_hydro}
\bibfield{author}{\bibinfo{person}{Cong Zhang}, \bibinfo{person}{J. Greenblatt}, \bibinfo{person}{M. Wei}, \bibinfo{person}{J. Eichman}, \bibinfo{person}{Samveg Saxena}, \bibinfo{person}{M. Muratori}, {and} \bibinfo{person}{Omar~J. Guerra}.} \bibinfo{year}{2020}\natexlab{}.
\newblock \showarticletitle{Flexible grid-based electrolysis hydrogen production for fuel cell vehicles reduces costs and greenhouse gas emissions}.
\newblock \bibinfo{journal}{\emph{Applied Energy}}  \bibinfo{volume}{278} (\bibinfo{year}{2020}), \bibinfo{pages}{115651}.
\newblock
\urldef\tempurl%
\url{https://doi.org/10.1016/j.apenergy.2020.115651}
\showDOI{\tempurl}


\bibitem[Zheng et~al\mbox{.}(2020)]%
        {chien2020}
\bibfield{author}{\bibinfo{person}{Jiajia Zheng}, \bibinfo{person}{Andrew~A. Chien}, {and} \bibinfo{person}{Sangwon Suh}.} \bibinfo{year}{2020}\natexlab{}.
\newblock \showarticletitle{Mitigating {Curtailment} and {Carbon} {Emissions} through {Load} {Migration} between {Data} {Centers}}.
\newblock \bibinfo{journal}{\emph{Joule}} \bibinfo{volume}{4}, \bibinfo{number}{10} (\bibinfo{year}{2020}), \bibinfo{pages}{2208--2222}.
\newblock


\bibitem[Zimmerman et~al\mbox{.}(2011)]%
        {matpower}
\bibfield{author}{\bibinfo{person}{R.~D. Zimmerman}, \bibinfo{person}{C.~E. Murillo-Sanchez}, {and} \bibinfo{person}{R.~J. Thomas}.} \bibinfo{year}{2011}\natexlab{}.
\newblock \showarticletitle{MATPOWER: Steady-State Operations, Planning and Analysis Tools for Power Systems Research and Education}. In \bibinfo{booktitle}{\emph{Power Systems, IEEE Transactions on, vol. 26}}. \bibinfo{pages}{12--19}.
\newblock


\end{thebibliography}

\end{document}